# Bononia, the Roman Bologna: Archaeoastronomy and Chronology

Amelia Carolina Sparavigna

Politecnico di Torino

In an article written by Giulio Magli on the orientation of the Roman towns, Bononia, the Roman Bologna, is given as a specific example to support Magli's thesis on the existence of preferred solar alignments of the urban layout. Assuming that the Roman towns had been oriented to the sunrise on a given day of the year, Magli suggested possible preferred alignments according to Roman festivals, in particular the festival of Terminalia. Of Bononia, we know the date of foundation as Roman colony in 189 BC, given by Livy. We will show that, according to Roman chronology and Republican calendar, it is impossible that Bononia had been oriented to the sunrise on the day given by Livy. The discrepancy is huge. Moreover, the direction of the decumanus cannot match the dates of Terminalia for 189 BC. However, if we consider that the layout that we see today is that of a recolonization under Octavian, we can have a certain agreement between the direction of the decumanus and the sunrise on the day of Terminalia (within three days), and a perfect agreement with the day of the festival of Armilustrium.  In the proposed discussion, we will stress in particular the problem of the discrepancy between the historical dates of Roman chronology and the Julian dates, those that we can find according to an astronomical analysis. This problem is general and must be properly considered in any archaeoastronomical analysis of Roman towns.

In [1], we find arXiv file of an article written by Giulio Magli on the astronomical orientation of a set of Roman towns [2]. The analysed towns have been chosen among the Roman towns in Italy. The aim of Magli's article was that of investigating a possible orientation of the main axis of the towns, the decumanus, according to the sunrise on a given day of the year, linked to Roman festivals, in the framework of a ritual inherited by Romans from Etruscans (Disciplina Etrusca).

In the abstract of [1], it is told that "As is well known, several Roman sources report on the existence of a town foundation ritual, inherited from the Etruscans, which allegedly included astronomical references". For what concerns Roman rituals and divination in general, Cicero is the main source. However, for the astronomical references in town foundation, let us stress that the Latin sources consist in a few passages in the literature of the Gromatici, the Roman surveyors [3]. This literature is concerning the subdivision of the land (centuriation) and is not specifically referring to the foundation of the towns, as observed in [4]. The above mentioned passages are not explaining the procedure in detail, but are simply telling that the direction of the decumanus was referring to the path of the sun and moon in the sky, and, sometimes, to the rising sun in the case of unpractised surveyors who mistook the sunrise direction with the East.

In the past, some scholars have considered these notes of the Roman surveyors as indicating a foundation of the colonies according to the sunrise, so that the decumanus had to be aligned along the direction of the sun rising on the natural horizon (see discussion in [5], a proposed example is the Roman Timgad [6], and see also Appendix). Cases of Roman towns that could have been astronomically oriented exist for sure; however we need to be very careful in drawing conclusions and the reason is the following. Usually, the Roman towns have a very good orientation "secundum naturam", that is, according to the nature of the site. Therefore their orientation can result only by chance as astronomical oriented according to the sky, "secundum coelum". The Roman Como is one of this cases: it seems oriented according to solstices (astronomical horizon), but its orientation is the only possible one according to the natural site of the town.

"As a first step" in his investigation concerning the orientation of the Roman towns in Italy, Magli analysed a set of 38 towns. The result of the analysis, according to the author, is that a non-random orientation patterns emerged. As stressed by Magli, the analysis had been made on a relatively small number of towns, without distinguish the periods of their foundation. Moreover, the horizon considered for the sunrise observation is the astronomical horizon and not the natural one. So, in the list given by Magli, we can have towns where we can find negligible or large differences between the sunrise on the astronomical plane and that on the natural horizon, because of the different local landscapes.

After a discussion of the sectors of the solar orientation, in [1], we find as a conclusion of the analysis the following. "The absence of towns in the sector between 19 and 29 degrees south of east is, of course, significant too … . Interestingly enough, it may be noticed that the solar dates corresponding to these azimuths locate in two periods which do not contain any relevant festivity of the Roman calendar, namely (very roughly, because the effective dates depend on the specific orientation and latitude) the second half of November and second half of January. This observation can be compared with the fact that, instead, dates falling between 10 and 19 degrees north or south of east may indicate important festivals of the Roman Calendar. In particular, in the second half of February (orientations south of east) many important festivals took place: the Parentalia … , the Lupercalia … , and the Terminalia, festival of the god Terminus, protector of the boundaries and of the city walls. It has been actually already proposed that the orientation of Bonomia (Bologna) was chosen in such a way that the sun was rising in alignment with the decumanus of the city on the day of Terminalia (Incerti 1999), and a fieldwork may lead to similar conclusions for other towns of this group as well." [1] I am quoting what we can read in arXiv [1], because I prefer avoiding any possible misunderstanding.

In his discussion, Magli is linking the solar orientations to Roman festivals. As a consequence, we have to investigate the link of the solar orientations to the Roman calendar. Let me stress that the link between festivals and orientation of centuriation is not mentioned by the Latin surveyors, so it appears in [1] as a proposal made by the author (see Appendix).

Here some comments in detail.

"The absence of towns in the sector between 19 and 29 degrees south of east is, of course, significant too" [1]. The "absence of towns" is not true, because we have Torino (Julia Augusta Taurinorum) and Julia Concordia Sagittaria for sure. In [1], Torino is given at 30

SE, but it is at 26°22' SE [7]. Julia Concordia [8] has the same direction of Torino (within plus/minus a degree).

For what concerns the sentence "it may be noticed that the solar dates corresponding to these azimuths locate in two periods which do not contain any relevant festivity of the Roman calendar …" [1], we have to note that the Roman festivals were linked to the Republican calendar, if we are considering the historical periods before 45 BC. In that year the calendar was reformed by Julius Caesar into a solar calendar. During the Republican period, the sun was not ruling the time in Rome. The manner the calendar was managed by the Roman priests during the Republic was totally different from the management of the Julian calendar, therefore we have not to imagine the Roman festivals as fixed as in our calendar. The republican calendar was roughly a lunisolar calendar, and, as we will see in the specific discussion concerning Bononia, we can have the calendar shifted from the seasons of several months (see [9] and references therein). Consequently, the observation that "the effective dates depend on the specific orientation and latitude", is not relevant, because large differences existed between the dates of the calendar and the Julian dates.

Magli continues his thesis stressing that, instead, "dates falling between 10 and 19 degrees north or south of east may indicate important festivals of the Roman Calendar" [1]. So we arrive to Bononia and Terminalia festivals. Of course we can repeat the observation previously done, but here we are interested in considering the specific case of this town, the Roman Bologna.

The layout of Bononia is properly discussed in [10]. The work [10] by Manuela Incerti is a remarkable analysis of the orientation according to the land where the town was founded. She is discussing the nature of the place and shows how the Roman surveyors operated to have the best results. About the orientation of the centuriation, she is telling that the courses of the existing waterways probably suggested placing the decumanus at right angle to waterways. In [10], it is also noticed that the planning mechanism of the colony is based on well-known geometric conditions, that is, based on the rectangular triangles which were supporting the surveyor's works. Since a direction of the decumanus of 102° 30' was determined, the author in [10] is mentioning a possible link to the sunrise on 23 February, a date which is given for the festivals of Terminalia too. In any case, she is stressing that, by quoting the astronomical date, she does not want to assert that Bononia was founded for magic and sacred reasons (this is clearly told in [10]). Moreover she is pointing out that it is necessary to make a conversion from Gregorian date to Roman date.

Let us note that the date of 23 February is the date that we have today for the sunrise in the direction of the decumanus. In the case of Bologna, the natural horizon is corresponding to the astronomical horizon. We can use software CalSKY, and find the sunrise azimuth corresponding to the direction of the decumanus. Let us note that CalSKY is giving sunrise and sunset azimuths on the astronomical horizon, and it is not considering the atmospheric refraction.

As we can see from the following screenshot, the best agreement is on February 24.

```
24 Feb 2019   Rise  :  7h01.2m   Set  : 17h55.2m   Transit: 12h27m52s
                     az=102.6°         az=257.6°            Altitude=36.1° Aqr
```

To consider the effect of the atmospheric refraction, we can use software Stellarium. If we imagine that the upper limb of the sun is observed, we obtain that the best agreement is on February 24 too. So, let us imagine that it is possible that the ancient surveyors observed the sun as soon as it was appearing above the horizon. However, this is not told in the Latin literature of the Gromatici.

In [1] and [10], we find mentioned the date for the foundation of the Roman colony of Bononia. It is 189 BC. So let us use software CalSKY for this year: we can see that the direction of the decumanus corresponded to the sunrise azimuth on 28 Feb 189 BC (Julian date).

```
28 Feb 189BC   Rise  :  6h53.7m  Set  : 17h47.8m  Transit: 12h20m22s
                      az=102.6°        az=257.6°           Altitude=36.1° Psc
```

If we repeat the analysis with Stellarium – atmospheric refraction and upper limb of the sun - we find again the date (Julian date) of 28 February. However, we need to stress that we have also another date, as we can see using CalSKY, which had the same sunrise azimuth.

```
20 Oct 189BC   Rise  :  6h25.0m  Set  : 17h19.1m  Transit: 11h52m24s
                      az=102.4°        az=257.4°           Altitude=36.0° Lib
```

Therefore, besides February 28 (Julian date), we have October 20 (Julian date) too.

Before continuing, let us stress that the dates we can obtain by means of astronomical analyses, known as the Julian dates, are totally different, for the ancient times, from the historical dates given by the Roman Republican Calendar. They are also different, before year 8 AD, from the dates of the Julian Calendar [11].

Here a table, after [12], for Roman year 189 BC to have the conversion of dates. K means Kalends. Quin is Quinctilis and Sext. Sextilis. The italic Roman numbers are giving the months in the Julian dates.

| K-Quin | K-Sext | K-Sept | K-Oct | K-Nov | K-Dec | - 190 |
|---|---|---|---|---|---|---|
| 4-iii-190 | 4-iv-190 | 3-v-190 | 1-vi-190 | 2-vii-190 | 31-vii-190 | |

| K-Jan | K-Feb | K-Mart | K-Apr | K-Mai | K-Iun | - 189 |
|---|---|---|---|---|---|---|
| 29-viii-190 | 27-ix-190 | 25-x-190 | 25-xi-190 | 24-xii-190 | 24-i-189 | |

| K-Quin | K-Sext | K-Sept | K-Oct | K-Nov | K-Dec | - 189 |
|---|---|---|---|---|---|---|
| 22-ii-189 | 25-iii-189 | 23-iv-189 | 22-v-189 | 22-vi-189 | 21-vii-189 | |

| K-Jan | K-Feb | K-Mart | K-Apr | K-Mai | K-Iun - 188 |
|---|---|---|---|---|---|
| 19-viii-189 | 17-ix-189 | 7-xi-189 | 8-xii-189 | 6-i-188 | 6-ii-188 |

Let us note that in 188 BC, according to [12], there was an intercalation of the Mercedonius, the intercalary month of the Roman republican calendar. From the table we have that February 28, 189 BC (Julian date) is corresponding to 7 Quinctilis, and October 20, 189 BC (Julian date), is corresponding to 11 Mercedonius, that is the above-mentioned intercalary month of Roman year 188 BC. In the case that we consider October 20, 190 BC, we have that it was corresponding to 24 Februarius, that is VI Kal. Mart., the Regifugium of Roman year 189 BC. Of course, if we assume an uncertainty of one day, we have that October 19, 190 BC, was VII Kal. Mart., the day of Terminalia (Terminalia happened the day before the Regifugium). In any case, this date is ten months **before** the foundation of the colony according to Livy. Therefore, we have to discharge it if we consider the year given by Livy as the time to get the works for the establishment of the colony started. To have a coincidence for Terminalia we have to arrive at year 178 BC. As a consequence, we have that in the Roman year 189 BC, the direction of the Decumanus was not matching the sunrise on the day of Terminalia for sure.

The date of foundation that we find in [1] and [10] are coming from Livy.

«Eodem anno ante tertium Kal. Ianuarias Bononiam Latinam coloniam ex senatus consulto L. Valerius Flaccus M. Atilius Seranus L. Valerius Tappo triumviri deduxerunt. Tria milia hominum sunt deducta; equitibus septuagena iugera, ceteris colonis quinquagena sunt data. Ager captus de Gallis Bois fuerat, Galli Tuscos expulerant.» (Livy, Ab urbe condita, XXXVII, 57, 7).

"In the same year, on the third day before the Kalends of January, a Latin colony, was established at Bononia by authorization of the Senate, by Lucius Valerius Flaccus, Marcus Atilius Serranus and Lucius Valerius Tappo, the board of three appointed for the purpose. Three thousand men were placed there; the cavalrymen received seventy iugera each, the rest of the colonists fifty each. The land had been taken from the Gallic Boii; the Boii had expelled the Etruscans". [Livy, with an English translation by Evan T. Sage, 1919, Cambridge, Harvard University Press and London, Heinemann, available at https://archive.org/details/livywithenglisht10livy/ ]

In the Roman Republican Calendar, December had 29 days. In the inclusive count used by Romans, "ante tertium Kal.", means 28 December. Actually, what was the corresponding Julian date that we find in Livy? We can use the tables given in [12] (but we can also arrive to the same conclusions if we use the Roman chronology discussed in [13]). Using [12], we see that the Kalends of January were in August, that is, in the month of August according to Julian Dates. 28 December 189 (Roman calendar) was on 17 August 189 BC (Julian date). The decumanus of Bononia was not aligned to the sunrise on the day mentioned by Livy, if we assume that the layout of the town at that time had the same orientation that we see toady. Of course, it exists the possibility that Bononia was subjected to a new foundation, or recolonization, which had changed the layout of the town too. We will discuss the case after some further considerations on year 189 BC.

Let us continue our archaeoastronomical analysis. Using CalSKY, we can see that the summer solstice was on June 26, 189 BC (Julian date). So we can tell that the colony of Bononia was, in origin, founded in the astronomical season of summer, whatever it means a "foundation" (see Appendix). However, at a first glance it seems also reasonable. The days are longer than in winter, so Romans used the natural light for the works; climate was warm, but not so hot, and therefore it was easier to perform the operations required by the foundation.

Let me stress that the results obtained by means of CalSKY and the Roman chronology are telling the following. The direction of the decumanus is not matching, according to the date given by Livy, the direction of the sunrise. The direction of the decumanus is not matching the sunrise azimuth on the day of Terminalia, according to the Roman Calendar in 189 BC. This result is obtained using the Tables in [12]. The above discussion is therefore showing that what Magli has proposed in [1] is not true for the case of Bononia, as we imagine it founded in 189 BC, with the specific direction of the decumanus that we see today.

As we have seen, huge differences exist between Roman calendar and Julian dates. These differences can be surprising, but we are sure of them because the Roman chronology of that period is based on the solar and lunar eclipses mentioned by Livy [9]. As told in [13], and discussed in [14], Livy did not make mistakes in reporting the dates. In any case, as we can see in [15], several scholars have discussed the Roman chronology, being almost unanimously in agreement with the chronology given in [15].

Let us consider another excerpt from [1], concerning the solar sectors. "On the northern side there are of course too few data to draw conclusions; however, the distribution between 9 and 25 degrees NE is at least intriguing: only five towns, concentrated in only two angles. The corresponding dates fall into the period 10-30 of April which, of course, includes the foundation of Rome (21 April)." Magli is also mentioning the Pantheon, referring to [16] and to the foundation of Rome on 21 April.

Just for curiosity: when was "April 21" in the year of the foundation of Bononia? Using the Tables on [12], we see that it was on 15 December 190 BC (Julian date).

Therefore, before drawing any conclusion concerning a possible link between the foundation of the Roman towns and the Roman festivals, it is necessary to investigate the Roman chronology of the towns very carefully. Let us stress that the Roman festivals were given according to the Roman calendar. In the case of many of the towns considered by Magli, it was the Republican Roman calendar. For what concerns the festivals, let us observe that also in the case of those which were "stativa", that is fixed to some days of the calendar, like the Ludi Apollinares from 208 BC [9], the historical dates of the festivals were moving and shifting in the Julian proleptic calendar. Two are the main reasons: 1) the Roman Republican calendar was not anchored to solstices, and 2) in this calendar the intercalation was of a month. The Terminalia were "stativa" for sure, because they were marking the moment in the Roman Calendar, where it was inserted the intercalation month of the Mercedonius, the additional "February", used to adjust the calendar. Moreover, the intercalation was often not properly applied. In spite of the fact that Terminalia were stativa, the corresponding Julian date was changing every year. This is the same as for the date of our Easter, which is calculated by means of a lunisolar calendar.

As we have told previously, there is the possibility that the layout of the Roman colony that we see today is not that of 189 BC. Actually, Incerti [10] is talking of an Imperial Bologna, so we need to understand if the layout of the town, the Roman town that we see today, is different from that originally planned in 189 BC.

We know that Bononia had recolonized under Antony and Octavian [17,18]. This is a conclusion coming from a rereading of what Pliny is telling in his Natural History about the towns in the northern part of Italy [19]. As told in [18], the city was involved, seemingly, without injury in the bellum Mutinense in 43 a.C.. The town was destroyed by fire in 53 AD and restored under Claudius. In [20], it is told that "the orthogonal urban plan created under Augustus is still discernible in the city today". However, when Augustus recolonized Bononia, had he changed the orientation of the decumanus?

A discussion in [21] tells the following. "Nell' interno della Bononia romana, presso la porta occidentale, un recentissimo rinvenimento ha mostrato che la via Emilia dirigevasi presso che parallelamente al primo tronco della via s. Felice (6), ma era un poco più a mezzodì della strada attuale, e si può supporre molto ragionevolmente che continuasse in linea retta quasi come l'attuale Mercato di mezzo fino a porta ravegnana, quindi attraversasse senza diversioni la città da occidente a oriente. Ne consegue per ciò che nello scavo d' una fossa lungo un tratto del Mercato di mezzo non poteva trovarsi come non si trovò il proseguimento dell' Emilia, che dovrebbe essere o essere stato sotto le case prospettanti a settentrione." That is, it seems that the Via Emilia, in origin, was crossing the town without diversions. Then, as far as I can argue, the decumani of the Imperial Bologna are probably different from those of the original colony, if they were parallel to the Via Emilia.

Of the new layout of Bononia we have not a date. Probably, Octavian recolonized it with his veterans after the Battle of Actium (2 September 31 BC). Then, let us consider year 31 BC. The direction of the decumanus corresponds to 27 February. As told in [11], 27 February (Julian date) is corresponding to 26 February in the Julian Calendar of the time. So we are three days after the Terminalia. Moreover, besides February 27 (Julian date), we have October 20 (Julian date), which is October 19 in the Julian Calendar. Actually, if we are not close to solstices, the direction of the decumanus is corresponding to two days, and not only to the day more convenient to the aim of the discussion.

```
27 Feb 31BC   Rise :  6h52.3m   Set :  17h48.5m   Transit: 12h20m06s
              az=102.2°         az=258.0°         Altitude=36.3° Psc
```

```
20 Oct 31BC   Rise :  6h25.6m   Set :  17h17.8m   Transit: 11h52m01s
              az=102.7°         az=257.1°         Altitude=35.8° Lib
```

On 19 October, in the Roman Calendar, we had the Armilustrium, a festival in honour of Mars, the god of war [22]. It was celebrated every year on the 14th before the Calends of November. This festival is reported in the Fasti Antiates Maiores, a painted wall-calendar from the late Roman Republic. It is archaeologically attesting the Roman calendar before the Julian calendar reform. Since the length of October was not changed by the reform of the calendar, the festival remained on 14th day before the Calends of November. On the day of the Armilustrium, the weapons of the soldiers were ritually purified and stored for winter. This festival was very important for the Octavian's veterans for sure. They were recolonizing Bononia, and in this town, they were storing the weapons for the rest of the life.

Therefore, if we consider that the layout that we are analysing is that of the recolonization under Octavian, we can have a certain agreement between the direction of the decumanus and Terminalia (within three days), and a perfect agreement for Armilustrium.

However, are these alignments significant? Or are they just coincidences? In any case, let me stress that, in the framework of solar orientations, Armilustrium is as important as Terminalia.

Let us conclude the discussion with the following observation. The discrepancy between the historical dates reported in the Roman republican calendar and the Julian dates that we can obtain from any astronomical analysis can be huge and changing from year to year. Moreover, the problem of the discrepancy is general, that is, not only linked to Bononia and its foundation. Therefore, the Roman chronology must always be properly considered in any archaeoastronomical analysis of Roman towns or buildings, in particular if we want to draw some conclusions about preferred solar orientations.

**Appendix**

Two observations are necessary. The first is concerning the foundation of colonies. It is a long sequence of works starting from the surveying of the area chosen for the colony and ending with the exhibition of the "forma urbis" in the forum [23]. In this long sequence of actions and days, what was the one that colonists commemorated? Different opinions exist as we can see in [24].

The second observation is concerning the direction of decumanus, coincident to a sunrise azimuth, and the link to Roman festivals. We can find it in Das Templum[1] by Heinrich Nissen [25]. The "Nissenschen Theorie" is also reported by Friedrich Nietzsche in his lesson about Der Gottesdienst der Griechen [26].

Here what Nissen is telling "Diese Erklärung, welche sich aus den Worten der Gromatiker mit Notwendigkeit ergiebt, eröffnet eine ganz neue Betrachtungsweise. Wie jeder Mensch, so hat auch der Gott und die Götterwohnung und das Templum in seinen verschiedenen Anwendungen überhaupt einen Geburtstag. Dies gilt ebenso von der Stadt: einige Geburtsjahre italischer Städte sind S. 56 zusammengestellt. So wenig wir hiervon wissen, erscheint unsere Kunde bezüglich der Geburtstage doch noch weit dürftiger. Für Rom wird er bezeichnet durch das Parilienfest am 21. April, für die Colonic Brundisium durch das Fest der Salus auf dem Quirinal am 5. August. Nach dem oben Gesagten muss also die Richtung des Decumanus entsprechen dem Sonnenaufgang am Gründungstag des Templum. Und um die Theorie auf gegebene Fälle anzuwenden, lässt sich aus dem Decumanus der Gründungstag finden, oder falls der Tag bekannt, die Richtung des Decumanus".

Nissen is telling that this explanation, which necessarily follows from the words of the gromatici, opens up an entirely new way of looking at these things. Like every human being, gods and god' temples, or the Templum in its various applications, they have a birthday in general. This also applies to the town: some years of birth of Italian cities are given S. 56. Little as we know about years then, and our sources seem even poorer when it comes to birthdays. For Rome it is designated by the festival of Parilien on April 21st, for the Colonic Brundisium by the festival of Salus on the Quirinal on August 5th. According to the discussion above, the direction of the decumanus must correspond to the sunrise on the day the Templum was founded. And to apply the theory to given cases, the founding day can be found in the Decumanus, or if we know the day, the direction of the Decumanus is given.

---

1  The Latin templum is the Greek temenos. It is a piece of land marked off from common uses and dedicated to a god, a sanctuary, holy grove or holy precinct.

Nissen's conclusions are coming from his interpretation of the literature of Roman surveyors and of Roman mythology. His approach had a consequence: the birthday of a colony was associated with the day when the direction of its decumanus was determined, but this is not true (see discussion in [24]).

As evidenced by Giulio de Petra in his review of Das Templum [27], Nissen was largely using the confirmation bias to determine the link of Templum to sunrise. De Petra's review should be read carefully, to acknowledge all the criticisms to weak points in Nissen's work. Regarding the orientation of decumani, sunrise and festivals, de Petra also stressed the problem of chronology, that is, even if we are considering Nissen's assumptions as reasonable, it is not clear how they can be useful in a practical manner, due to the difficulty of manage the chronological problems.

A discussion concerning Brindisi in given in [28] (in Italian).